\pdfoutput=1
\documentclass[aps,pre,reprint,superscriptaddress,nofootinbib,a4paper]{revtex4-2}
\usepackage[utf8]{inputenc}
\usepackage[T1]{fontenc}
\usepackage{amsmath,amssymb}
\usepackage{graphicx}
\usepackage{eso-pic,pgffor,refcount}
\usepackage{xcolor}
\usepackage[nice]{nicefrac}
\usepackage{lmodern}
\usepackage{float}
\usepackage[colorlinks=true,allcolors=blue]{hyperref}

\setcitestyle{super,sort&compress}
\makeatletter
\renewcommand{\figurename}{Fig.}
\renewcommand{\fnum@figure}{\textbf{\figurename~\thefigure}}
\def\@caption@fignum@sep{\ \textbf{\textbar}\ }
\def\@hangfrom@section#1#2#3{\@hangfrom{#1#2}#3}
\def\section{\@startsection{section}{1}{\z@}{0.9cm \@plus1ex \@minus .2ex}{0.35cm}{\normalfont\large\bfseries\raggedright}}
\def\subsection{\@startsection{subsection}{2}{\z@}{.6cm \@plus1ex \@minus .2ex}{0.2cm}{\normalfont\normalsize\bfseries\raggedright}}
\makeatother

\newif\ifshowtodos
\showtodostrue
\ifshowtodos
  \newcommand{\todo}[1]{\textcolor{red}{\small[\textbf{TO COMPLETE:} #1]}}
  \newcommand{\fillin}[1]{\textcolor{red}{\ifmmode\langle #1\rangle\else$\langle$\textbf{#1}$\rangle$\fi}}
\else
  \newcommand{\todo}[1]{}
  \newcommand{\fillin}[1]{\textcolor{red}{\textbf{??}}}
\fi

\begin{document}

\title{Resolving non-neutrality across scales in a hyperdiverse tropical forest}

\author{Pietro Zgaga}
\affiliation{Networks Unit, IMT School for Advanced Studies, Lucca, Italy}
\author{Pablo Villegas}
\email{pablo.villegas@cref.it}
\affiliation{`Enrico Fermi' Research Center (CREF), Via Panisperna 89A, 00184 Rome, Italy}
\affiliation{Instituto Carlos I de F\'isica Te\'orica y Computacional, Universidad de Granada, E-18071 Granada, Spain}
\author{Tancredi Caruso}
\affiliation{School of Biology and Environmental Science, University College Dublin, Dublin, Ireland}
\author{Tommaso Gili}
\affiliation{Networks Unit, IMT School for Advanced Studies, Lucca, Italy}

\begin{abstract}
Different ecological processes operate at different spatial and temporal scales. Macroscopic regularities can therefore be captured by descriptions that ignore microscopic details, even when those details are mechanistically relevant to ecological processes. Here, we develop a framework inspired by statistical mechanics to track interspecific spatial correlations under spatial coarse-graining. To do so we combine random-matrix filtering with a spatially explicit neutral model. Applying this approach to eight censuses of the Barro Colorado Island tropical forest, we observe a continuous crossover from an interacting to an effectively neutral organization. Below a few tens of metres (20--35~m), the empirical correlation spectrum differs from neutral expectations, revealing scale-dependent species-specific structure. Part of this structure can be attributed to environmental heterogeneity. However, residual spatial associations provide candidate signatures of biotic interactions or other unmeasured ecological, non-neutral processes. With increasing spatial coarse-graining, empirical and neutral correlation spectra progressively converge, and above the crossover scale the neutral model provides an accurate description of the measured collective organization. Neutrality thus emerges as an effective description of the collective organization of the forest above a characteristic scale, without implying that species are equivalent at the individual level. Our framework provides an operational way to locate that scale.
\end{abstract}

\maketitle

Ecological communities display striking regularities in biodiversity patterns despite the immense diversity of species and interactions they contain \cite{Magurran2004, Vellend2016}. A central question in community ecology is how descriptions of the same system at different scales relate to one another \cite{Leibold2004, Adler2007, Thompson2020}. Neutral theory has proven remarkably successful in reproducing several macroecological patterns observed in tropical forests, including species–abundance distributions, species–area relationships, and spatial clustering of individuals\cite{Hubbell2001, Volkov2003, Plotkin2000, Azaele2016, ODwyer2018}. At the same time, abundant empirical evidence shows that individuals belonging to different species differ substantially in traits, environmental preferences, and competitive abilities\cite{Harms2001, John2007, Clark2018}, suggesting that ecological interactions and environmental filtering play a fundamental role in shaping community structure even when, at some level of approximation, some processes may approach neutrality in the limit of high species richness \cite{Wiegand2012}.

Reconciling the predictive success of neutral models with the evident reality of species differences has been a notorious issue in ecology\cite{Chave2004, Tilman2004, Rosindell2011}, partly owing to an implicit assumption, which is that these theories represent mutually exclusive interpretations of the nature of ecological interactions. However, the adequacy of stochastic processes in synthesizing macroscopic patterns does not imply the absence of strong differentiation and interactions among individuals. Neutrality may instead be an effective description, valid for the collective organization of a community whose individuals are far from equivalent. In particular, it has been proposed that neutral processes may emerge as an effective large-scale statistical description of ecosystems once the complex web of species interactions is coarse-grained across sufficiently large spatial or temporal scales\cite{Chave2004, Fisher2014, Azaele2016}, and that the relative importance of species-specific and stochastic processes is itself scale dependent\cite{Shmida1985, Chase2011}. Identifying the spatial scale above which a neutral description becomes sufficient is, however, still an open problem.

Spatial structure provides a natural window into this question. Tropical forests exhibit strong spatial organization, including clustering of individuals, large density fluctuations, and scale-dependent vegetation patterns\cite{Condit2000, Plotkin2000, Villegas}. These spatial signatures indicate that the processes shaping biodiversity leave measurable imprints in the spatial arrangement of communities. While the spatial organization of individual species has been extensively studied using point-pattern statistics such as the pair correlation function\cite{MoloneyPP, Detto2013, Villegas}, much less is known about the correlation structure between different species within the same community. Yet interspecific spatial correlations provide direct information about the processes governing species coexistence, including dispersal, environmental filtering, and ecological interactions\cite{MullerLandau2008, Volkov2009, Baldeck2013a}.

Spatially explicit neutral models can reproduce several qualitative features of tropical forests, including clustered point patterns and realistic abundance fluctuations\cite{ChaveLeigh2002, Azaele2016, Pigolotti2018}, suggesting that neutral demographic dynamics capture an important component of community organization. At the same time, recent studies have reported scale-invariant vegetation clusters and anomalous density fluctuations in forests such as Barro Colorado Island\cite{Villegas, Detto2016}, indicating that ecological processes in tropical rainforests may involve multiple intertwined spatial scales. These observations raise a fundamental question: starting from individuals and species interacting at local scales, does the resulting collective organization of ecological communities become effectively neutral at large scales?

Here we address this question by developing a statistical framework to quantify interspecific spatial correlations across scales. Our approach combines spatial coarse-graining, random-matrix filtering of sampling noise, and comparison with a spatially explicit neutral model, in the spirit of coarse-graining analyses of collective behavior in other biological systems\cite{Meshulam2019, LRG}. Applying this framework to the long-term forest census of Barro Colorado Island, we identify a characteristic crossover scale above which the empirical correlation structure becomes statistically compatible with neutral expectations. Below this scale, non-neutral structure is clearly detectable and can be further decomposed into environmentally mediated and residual spatial associations.

Our results support a scale-dependent view of ecological organization: species-specific and environmental structure is detectable locally, whereas spatial coarse-graining progressively renders these microscopic details irrelevant to the measured collective correlations, so that a neutral description becomes sufficient at large scales. Neutrality, in this view, is an emergent property of the collective organization of the forest.

\begin{figure*}[t]
 \centering
 \includegraphics[width=\textwidth]{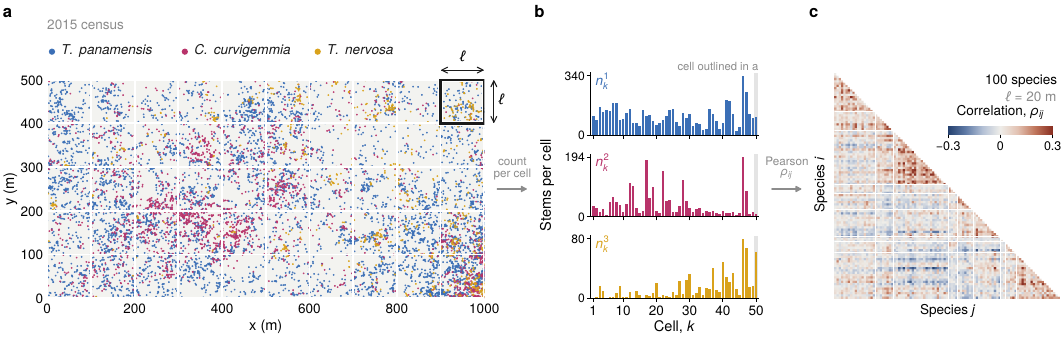}
 \caption{
\textbf{Interspecific spatial correlations across scales.}
\textbf{a}, Spatial distribution of three representative species, \textit{Tetragastris panamensis}, \textit{Coussarea curvigemmia} and \textit{Talisia nervosa}, in the Barro Colorado Island (BCI) forest plot in the 2015 census. The plot is partitioned into cells of side $\ell=100$~m (white grid) to illustrate how point-pattern data are coarse-grained into spatial abundance fields; one cell is outlined.
\textbf{b}, Spatial abundance series obtained from the same partitioning, each species being represented by its number of stems in each cell, $n^i_k$; grey band, the cell outlined in \textbf{a}. This transformation is what allows the spatial co-occurrence structure of the community to be analyzed through correlation matrices at any scale.
\textbf{c}, Pearson correlation matrix between the 100 most abundant species at $\ell=20$~m (mean of the eight censuses; the matrix is symmetric and only its lower triangle is shown). Species are ordered by hierarchical clustering of the noise-filtered matrix (white lines separate the clusters), so that groups of co-occurring species appear as blocks along the diagonal.
}
 \label{fig:framework}
\end{figure*}

\section*{Results}
\subsection*{Interspecific spatial correlations across scales}
To investigate how interspecific spatial associations vary across scales, we analyzed the long-term forest census of Barro Colorado Island (BCI), one of the most extensively monitored tropical forest ecosystems\cite{Leigh1999, Condit2000, Legendre2019}. The dataset comprises eight censuses spanning several decades, recording the precise location of more than $10^5$ trees and shrubs belonging to roughly 300 species within a 50-ha forest plot. This exceptional spatial resolution allows the community to be analyzed as a spatial point process, enabling a direct investigation of how interspecific correlations emerge across scales.

We partitioned the forest plot into spatial cells of side length $\ell$ and quantified the abundance of each species within each cell. This spatial coarse-graining transforms the two-dimensional distribution of individuals into a set of spatial abundance series for each species (Fig.~\ref{fig:framework}a,b). From these series, we computed the pairwise correlation matrix between species, which captures the tendency of species to co-occur or spatially exclude one another at a given spatial scale. To ensure sufficient sampling statistics\cite{PRX_Gar,Izaro2026}, the analysis focuses on the 100 most abundant species in the plot, each represented by at least $\sim 200$ individuals (Supplementary Note 1).

The resulting correlation matrices display a complex structure containing both positive and negative correlations (Fig.~\ref{fig:framework}c). However, raw correlations inevitably include spurious contributions arising from finite sampling and demographic stochasticity. To isolate meaningful ecological structure, we therefore analyzed the eigenvalue spectrum of these correlation matrices.

\subsection*{Random-matrix filtering of sampling noise}
The eigenvalue spectrum of the empirical correlation matrix reveals a clear separation between noise-dominated and structure-dominated components. Random matrix theory, long used to characterize the structure and stability of ecological communities\cite{Allesina2012, Barbier2018}, predicts that purely random correlations generate an eigenvalue distribution following the Marchenko–Pastur law\cite{Marchenko1967, Plerou1999, Laloux1999} (see Methods), allowing statistically meaningful correlations to be distinguished from fluctuations arising from finite sampling.

When comparing the empirical spectrum with the theoretical prediction, we observe that most eigenvalues fall within the expected noise interval, while a smaller set of eigenvalues lies significantly outside it (Fig.~\ref{fig:panel15}c). These outlying eigenvalues correspond to genuine ecological structure embedded in the spatial organization of species.

By filtering out the noise-dominated portion of the spectrum, we obtain a denoised correlation matrix that retains only statistically significant correlations between species. Once the species are reordered according to their correlation structure (Fig.~\ref{fig:framework}c; raw and filtered matrices are compared in Supplementary Note 2), this matrix reveals a block organization, with clear patterns of positive and negative associations that are otherwise obscured by sampling noise.

\subsection*{A neutral description above a characteristic scale}
\begin{figure*}[t]
    \centering
    \includegraphics[width=\textwidth]{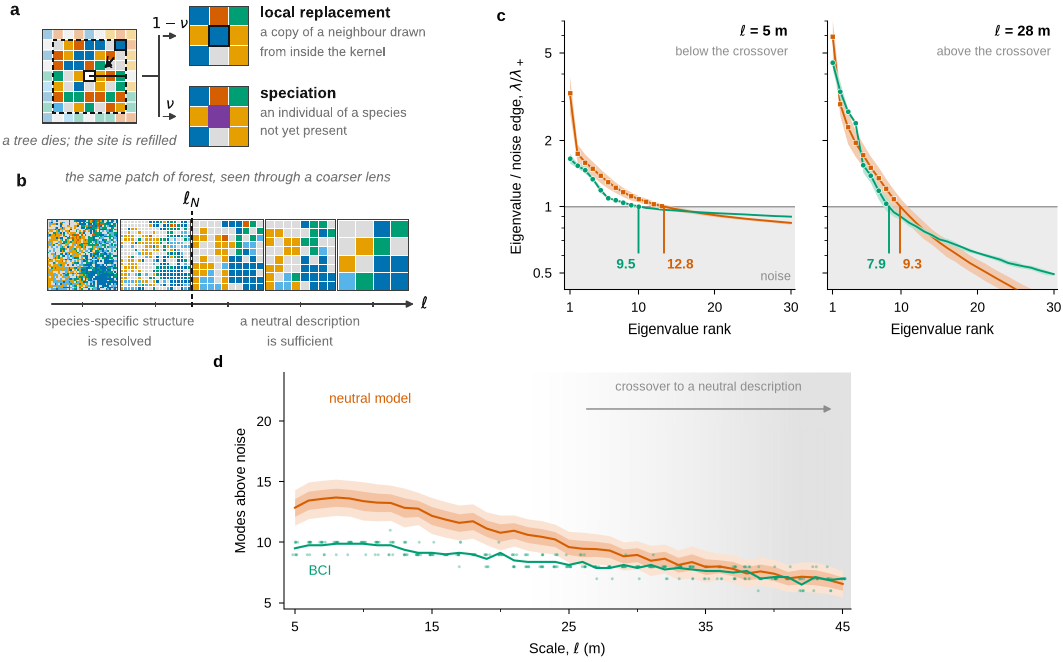}
    \caption{\textbf{Identifying the spatial scale of neutrality in ecological communities.}
  \textbf{a}, Schematic representation of neutral dynamics: a tree is removed at random and gets replaced either by an individual belonging to a species occurring within radius (\textit{kernel}) \textit{k} or, with probability $\nu$, by a new species. \textbf{b}, As the system is analyzed with a coarser lens, species-specific structure becomes irrelevant and a neutral description of its spatial correlations becomes sufficient. \textbf{c}, Leading eigenvalues of the species--species correlation matrix at $\ell=5$~m (left, below the crossover) and $\ell=28$~m (right, above it), for BCI (green; line, mean of the eight censuses; band, their range) and for the spatially explicit neutral model (SENM; orange; mean and two standard deviations over 100 realizations). Each spectrum is divided by the Marchenko--Pastur edge $\lambda_+$ of its own data set, which differs between the two because the two domains contain different numbers of cells, so that the region compatible with sampling noise ($\lambda/\lambda_+<1$, grey) is the same for both; ranks beyond 30 lie in it and are not shown. Markers, eigenvalues above the edge; ticks and numbers, mean number of such modes.
\textbf{d}, Number of non-noise eigenvalues as a function of the scale $\ell$, for BCI (dots, individual censuses; line, mean) and for the SENM (mean, with bands of one and two standard deviations). At small scales the neutral model carries many more non-noise modes than the empirical community; as $\ell$ grows the two curves converge. The grey shading marks the crossover: from about 25~m the BCI mean enters the two-standard-deviation band of the neutral ensemble, and at larger scales the two curves merge, so that a neutral description of this observable becomes sufficient.}
    \label{fig:panel15}
\end{figure*}

Neutral demographic dynamics can by itself generate structured interspecific correlations through dispersal and stochastic birth–death processes, even in the absence of species-specific interactions\cite{ChaveLeigh2002, Rosindell2011}. To determine whether the empirical structure can be accounted for by such dynamics, we compared the correlation spectra measured in BCI with those generated by a spatially explicit neutral model calibrated to the same system (see Methods). At small spatial scales, the empirical and neutral spectra are clearly distinct, indicating that neutral dynamics alone does not reproduce the observed correlation structure. As the spatial scale increases, however, the two spectra progressively converge, apart from their two or three leading modes, discussed below (Fig.~\ref{fig:panel15}c). The same crossover is apparent in the scale dependence of the number of non-noise eigenvalues (Fig.~\ref{fig:panel15}d). The convergence comes from the neutral side: the number of non-noise modes in BCI changes little with scale, from about ten at 5~m to seven at 45~m, whereas the neutral model, whose short dispersal kernel produces many correlated modes at fine resolution, loses them as $\ell$ grows. At sufficiently large scales, the number of non-noise modes of the empirical spectrum falls within the variability of the neutral ensemble, indicating that the neutral model provides an effective description of the measured interspecific correlation structure.

This crossover defines a characteristic spatial length, which we term the neutrality scale, \(\ell_N\). It is not a sharp threshold but the range of scales over which the number of non-noise modes of the empirical spectrum moves into the variability of the neutral ensemble: the BCI mean enters the two-standard-deviation band of the neutral model at about 25~m, and at larger scales the two curves merge (Fig.~\ref{fig:panel15}d). In the BCI forest, $\ell_N\approx25$--30~m, within the 20--35~m window in which the conspecific pair probability also crosses over to its neutral form (Fig.~\ref{fig:pairs}). Above \(\ell_N\), neutrality should therefore be interpreted as an effective coarse-grained description rather than as a statement of microscopic ecological equivalence. Above $\ell_N$ the agreement concerns the number of collective modes and the variance they carry, not every mode: the two or three leading eigenvalues of BCI remain larger than in the neutral ensemble (Fig.~\ref{fig:panel15}c) and grow with scale.

Below the neutrality scale, the filtered correlation matrix contains both positive and negative interspecific associations. Positive correlations indicate spatial co-occurrence, whereas negative correlations indicate spatial segregation. These patterns may arise from several mechanisms, including shared or contrasting environmental preferences, dispersal processes, facilitation, or competition\cite{Harms2001, John2007, Volkov2009}. Importantly, many correlations present in the raw matrix vanish after random-matrix filtering, showing that a substantial fraction of apparent associations is attributable to sampling noise.

\subsection*{The same crossover in the conspecific pair probability}
\begin{figure*}[t]
 \centering
 \includegraphics[width=\textwidth]{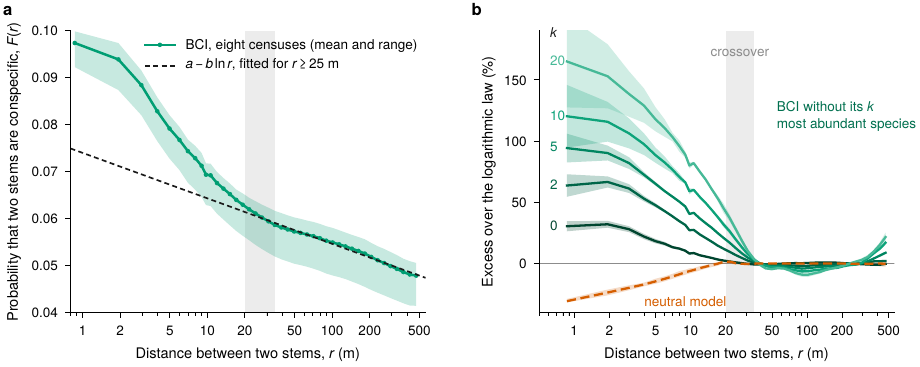}
 \caption{\textbf{The same crossover in the conspecific pair probability.}
\textbf{a}, Probability $F(r)$ that two stems a distance $r$ apart belong to the same species (all stems $\geq$1~cm dbh; mean and range of the eight censuses). Dashed line, logarithmic law $a-b\ln r$ fitted for $r\geq25$~m and extrapolated to shorter distances.
\textbf{b}, Relative excess over the logarithmic law, $F(r)/(a-b\ln r)-1$, each curve with its own fit for $r\geq25$~m. Green, BCI with the $k$ most abundant species removed (species with at least 200 stems; mean and range of censuses 4 and 8). Orange, spatially explicit neutral model calibrated on the slope of $F(r)$ at large distances (mean and range of three realizations; see Methods). Grey band, 20--35~m.}
 \label{fig:pairs}
\end{figure*}

The crossover is not specific to the correlation spectrum. A second, independent observable is the probability $F(r)$ that two stems a distance $r$ apart belong to the same species, the classical spatial statistic of neutral theory, for which spatially explicit neutral models predict a logarithmic decay with distance in two dimensions\cite{ChaveLeigh2002, Condit2002}. Unlike the correlation matrices, $F(r)$ is built from the distances between individual stems rather than from counts in cells. In BCI, $F(r)=a-b\ln r$ holds from 25 to 500~m in each of the eight censuses ($R^2=0.98$--$0.996$), with a slope that does not change over 33 years (Fig.~\ref{fig:pairs}a). At shorter distances the data rise above the logarithmic law, and conspecific pairs 5~m apart are 15--21\% more frequent than the law predicts.

Because $F(r)$ weighs each species by the square of its abundance, this excess could in principle be carried by the two dominant species alone. It is not: removing the 2, 5, 10 or 20 most abundant species makes the short-range excess larger, and every curve rejoins the logarithmic law between 20 and 35~m (Fig.~\ref{fig:pairs}b). The same crossover is found separately in saplings and in trees above 10~cm dbh, which are different individuals, and in most species taken one at a time (Supplementary Note 4). A spatially explicit neutral model calibrated on the large-scale slope of $F(r)$ (see Methods) follows the logarithmic law down to $\sim$15~m and falls below it at shorter distances, the opposite of the forest. Above the crossover, the forest and a model without species-specific interactions share the same functional form; below it, they part ways. Two observables constructed in different ways therefore locate the crossover in the same window of scales.

Having identified the spatial regime in which the empirical community departs from neutral expectations, we next asked which components of this non-neutral structure could be explained by environmental heterogeneity within the forest plot.

\subsection*{Environmental filtering of residual correlations}
Residual correlations between species may arise not only from direct ecological interactions, but also from shared responses to environmental gradients\cite{Harms2001, Baldeck2013a, Baldeck2013b}. To assess the role of environmental filtering, we compared the filtered correlation structure with spatial maps of soil properties across the plot\cite{John2007, Dalling2012, Condit2013}.

\begin{figure*}[t]
 \centering
 \includegraphics[width=\textwidth]{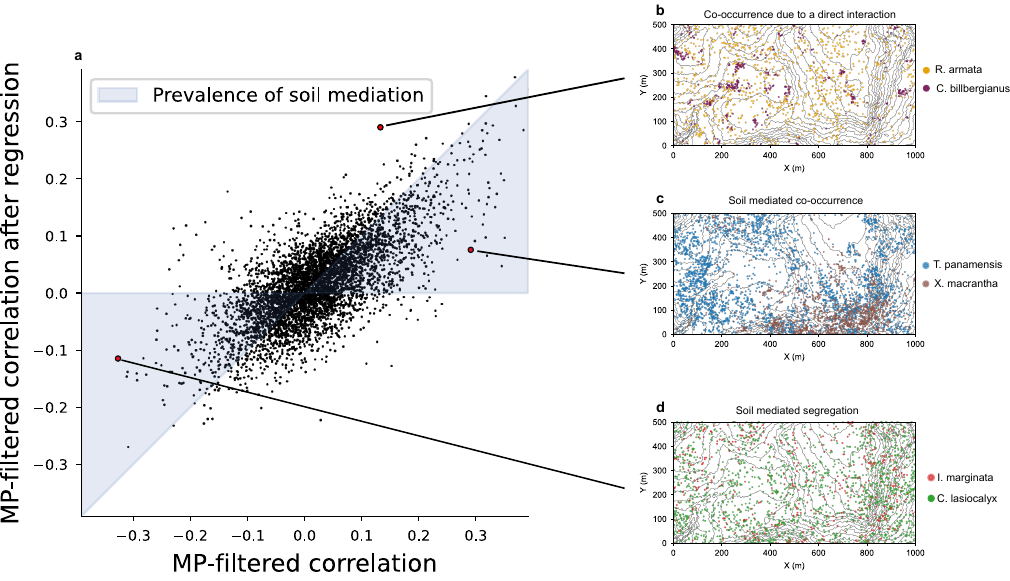}
\caption{
\textbf{Environmental filtering contributes to interspecific spatial correlations.}
\textbf{a}, Filtered pairwise correlation between species before and after the soil-driven component has been removed by environmental regression; each point is one species pair. Points on the diagonal are left untouched by the regression, and their association is independent of the measured soil variables. The shaded wedge, bounded by the diagonal and the horizontal axis, collects the pairs whose correlation is driven toward zero without changing sign, that is, those for which the environmental and the residual contributions reinforce each other and soil mediation prevails. Points lying above the diagonal are the opposite case: there the two contributions oppose each other, the raw correlation understates the underlying association, and removing the soil component makes the correlation stronger. The three pairs highlighted in red are the ones shown in \textbf{b}--\textbf{d}.
\textbf{b}, \textit{R.~armata} and \textit{C.~billbergianus}, a pair whose association is masked by the soil gradients and emerges only after regression.
\textbf{c}, \textit{T.~panamensis} and \textit{X.~macrantha}, whose co-occurrence is largely accounted for by a shared environmental preference.
\textbf{d}, \textit{I.~marginata} and \textit{C.~lasiocalyx}, segregated in space by contrasting soil preferences rather than by any direct antagonism. Contour lines in \textbf{b}--\textbf{d} indicate the topography of the plot.
}
 \label{fig:soil}
\end{figure*}

After accounting for correlations explained by soil variables, such as nutrient concentrations and pH\cite{John2007, Baldeck2013a, Condit2013}, several species pairs previously appearing positively correlated exhibit weaker or vanishing associations (Fig.~\ref{fig:soil}a). This analysis indicates that part of the residual correlation structure is mediated by environmental factors, while other correlations remain unexplained by soil variables and may therefore reflect direct ecological interactions.

The two axes of Fig.~\ref{fig:soil}a make this decomposition explicit. A filtered correlation can be written as the sum of a component induced by the response of both species to the measured soil gradients and a residual component that those gradients do not explain. When the two contributions carry the same sign they reinforce each other, so removing the environmental part necessarily drives the correlation toward zero without changing its sign: these pairs populate the shaded wedge between the diagonal and the horizontal axis, which we therefore take as the region where soil mediation prevails. Pairs falling on the diagonal are unaffected by the regression, and their association is independent of the measured environment.

Pairs lying \emph{outside} the wedge, above the diagonal, behave in the opposite way, and they are the most informative of the three cases. For these the environmental and residual contributions must act in opposite directions and partially cancel, so that the raw correlation understates the underlying association and the regression makes it stronger rather than weaker. \textit{R.~armata} and \textit{C.~billbergianus} are of this kind (Fig.~\ref{fig:soil}b): the soil gradients tend to pull the two species apart, while some other process keeps them together, and the association only becomes visible once the environmental contribution has been removed. Such pairs would be missed entirely by an analysis based on raw correlations alone.

In contrast, other species pairs that initially appear strongly correlated exhibit a substantial reduction of correlation after accounting for soil heterogeneity (Fig.~\ref{fig:soil}c), as in the case of \textit{T.~panamensis} and \textit{X.~macrantha}. The spatial distributions of these species reveal that their co-occurrence is largely driven by shared environmental preferences rather than direct interactions. 

Finally, some species pairs exhibit spatial segregation. In some cases, the species occupy distinct environmental niches associated with different soil conditions, leading to apparent anticorrelations that arise from habitat filtering rather than competitive exclusion, as for \textit{I.~marginata} and \textit{C.~lasiocalyx} (Fig.~\ref{fig:soil}d). In other cases, a negative correlation persists even after accounting for soil effects, suggesting the presence of an antagonistic interaction.

After accounting for measured environmental heterogeneity, a subset of spatial associations remains unexplained. These persistent correlations provide candidate signatures of biotic interactions, although contributions from unmeasured environmental variables or other spatial processes cannot be excluded.

\subsection*{Candidate biotic interactions}
Beyond identifying environmentally mediated correlations, our framework reveals species pairs whose spatial associations persist after filtering noise-dominated modes and accounting for measured environmental gradients. Because these associations occur within the spatial regime where the empirical community departs from neutral expectations, they provide statistically grounded candidates for non-neutral biotic structure\cite{Volkov2009}. Positive residual associations may reflect facilitation, shared ecological strategies, or unmeasured environmental drivers, whereas negative associations may reflect competition, contrasting ecological strategies, or other mechanisms of spatial segregation.

Inspection of specific species pairs highlights different ecological scenarios. In some cases, species exhibit robust positive spatial correlations that remain largely unchanged after all filtering steps, suggesting persistent co-occurrence that may arise from facilitative interactions, shared dispersal limitation, or unobserved environmental drivers\cite{MullerLandau2008, Clark2018}. In other cases, strong negative correlations persist even after environmental regression, indicating spatial segregation consistent with competitive interactions or contrasting ecological strategies.

These examples illustrate how the combination of spectral filtering and environmental regression progressively narrows the set of observed spatial associations. The correlations that remain are not, by themselves, evidence of direct causal interactions; rather, they provide a set of empirically grounded hypotheses for species pairs whose co-occurrence or segregation cannot be accounted for by sampling noise or the measured environmental variables.

As an example, we take the pair with the largest residual correlation ($+0.376$) after the filtering pipeline: \textit{L.~platypus}--\textit{H.~triandra}. The former is a large canopy tree that produces big, pulpy fruits, which ripen on the ground and are usually consumed by small mammals. The latter is an understory shrub that favors indirect sunlight and produces small, berry-like fruits often consumed by birds and arboreal monkeys. This complementarity between traits provides ecological plausibility for the residual association identified by our method, while not establishing a direct causal interaction.

\section*{Discussion}

Our results support the view that neutrality is an emergent property rather than an assumption about individuals and species: a neutral description becomes sufficient for the collective organization of the community even though its dynamics is shaped by a large number of species-specific interactions. The measured crossover makes this claim quantitative: neutrality is not an assumption to be accepted or rejected for the community as a whole, but a property that can be granted or denied to a given observable at a given resolution.

Importantly, effective neutrality does not imply that ecological interactions or environmental heterogeneity are absent above the crossover scale; the leading modes of the forest remain stronger than neutral there. Rather, they become irrelevant under coarse-graining for the particular collective observable considered here: the spectrum of interspecific spatial correlations. In this sense, there is a transition between a regime in which microscopic ecological structure is detectable and a regime in which a simpler macroscopic description is sufficient. The
crossover is also not a property of one statistic: two observables built in
different ways, one from counts in cells and one from distances between
individual stems, depart from the neutral form below a few tens of metres
and follow it above (Figs.~\ref{fig:panel15} and~\ref{fig:pairs}).

Below this crossover, our framework provides a complementary route for identifying the sources of non-neutral structure. Removing noise-dominated correlation modes and accounting for measured environmental heterogeneity leaves a set of residual interspecific associations that can be used to formulate hypotheses about candidate biotic interactions. Such residual associations should not be interpreted as direct causal evidence, because unmeasured environmental or spatial processes may contribute to them.

What sets the value of $\ell_N$ remains an open question. The crossover we
measure is of the same order as several characteristic lengths of this
forest: the crown diameter of a canopy tree, the distance over which
conspecific clustering decays in the plot\cite{Condit2000, Villegas}
(Fig.~\ref{fig:pairs}), and the bulk of the seed-dispersal kernel of most
species\cite{MullerLandau2008}. Two neutral estimates point the same way.
Fitting the neutral species--abundance distribution to this plot, Hubbell
found that about one tree in ten is an immigrant, the fraction of the plot
that lies within 17~m of its boundary\cite{Hubbell2001}; and the neutral model
that reproduces the large-scale decay of $F(r)$ has a dispersal kernel of
half-width $\sim$20~m (Methods).
This coincidence suggests that $\ell_N$ marks the scale at which cells become
large enough to average over the neighborhood within which an individual
competes and disperses, so that the identity of a tree's neighbors stops
constraining the occupancy of a cell. In species-rich forests this is
reinforced by recruitment limitation: many sites are won by whichever
species happens to arrive rather than by the best competitor for
them\cite{Hurtt1995, Hubbell1999}. If this interpretation is correct,
$\ell_N$ should track dispersal distance rather than species richness, a
prediction that can be tested directly by applying the same analysis to
forests with contrasting dispersal syndromes.

Viewed as a physical procedure, the analysis is a real-space coarse-graining
in which the neutral ensemble plays the role of the reference description and
the correlation spectrum plays the role of the observable followed along the
flow\cite{Meshulam2019, LRG}. We stress that this is an analogy with a
limited scope: we do not claim a scale-invariant fixed point, and the
approach to neutrality we observe is a crossover within a bounded window of
scales rather than a critical phenomenon. What the analogy does buy is a
criterion. Rather than asking whether species are equivalent, which is a
statement about individuals that spatial data cannot settle, one asks at
which resolution a description that treats them as equivalent stops being
distinguishable from the data. Hubbell argued that the success of neutral
theory implies community dynamics of much lower dimensionality than species
richness would suggest, because niche differences among BCI trees are
arrayed along a few life-history trade-offs shared by all species, and
because species that gain in some places and lose in others must have net
growth rates close to zero over large areas\cite{Hubbell2001}. The handful
of non-noise modes we find among one hundred species is a direct measurement
of that low dimensionality, and spatial coarse-graining is the averaging over
which local gains and losses cancel.

Several limitations bound these conclusions. Compatibility with the neutral
ensemble is established for one collective observable, and agreement in the
correlation spectrum does not imply that every property of the community
becomes neutral above $\ell_N$; observables sensitive to higher-order or
temporal structure may well retain a signature of species differences at
larger scales\cite{Detto2016}. Species-resolved statistics, such as which
pairs of species are spatially associated, retain habitat-driven structure
well above 30~m in this plot\cite{Wiegand2012}, and spatially explicit
neutral models reproduce several of its community patterns one at a time but
not all of them jointly\cite{May2015}. The comparison is also conditional on the
calibration of the neutral model, and on restricting the analysis to the 100
most abundant species, so the behavior of rare species is not probed. Finally,
the 50-ha plot bounds the range of accessible scales from above, and the
census interval leaves the temporal dimension of the problem untouched: the
eight censuses are treated here as independent realizations rather than as a
trajectory.

These limitations also indicate where the framework should be taken next.
Because it requires only mapped positions and a null model, the same
measurement can be repeated across the network of large forest plots, which
would show whether a neutrality scale exists generally and how it varies with
forest type, diversity, and disturbance regime\cite{Condit2002, Legendre2019}.
More broadly, nothing in the construction is specific to trees. Any system
whose components are sampled repeatedly in space admits the same question,
and the crossover scale, where it exists, provides an operational answer to
how coarse a description has to be before its constituents can be treated as
interchangeable.

\section*{Methods}
\subsection*{Data}
We used the eight censuses of the 50-ha Forest Dynamics plot on Barro Colorado
Island, Panama\cite{Condit2019}, a $1000\times500$~m rectangle in
which every free-standing woody stem above 1~cm diameter at breast height is
mapped, tagged, measured and identified to species, with censuses roughly every
five years.

\subsection*{Pearson correlation matrix}
We conceived a method to identify the interspecific spatial correlations at a distance $\ell$ (in m), building the correlation matrix between the different species. The method follows the next steps: (i) The study area is divided into different square boxes of side $\ell$, forming a grid of $T=n_x\cdot n_y$ elements. (ii) Boxes are put in sequential order, generating different occupation vectors accounting for the total number of trees of the $i$th species, $n^i_k=(n^i_1,n^i_2,...,n^i_T)$. (iii) Compute the Pearson's correlation coefficient, i.e., the correlation between vectors $n^i$ and $n^j$ over all species,

\begin{equation}
r_{ij}=\frac{\underset{k}{\sum}n_{k}^{i}n_{k}^{j}-T\left\langle n^{i}\right\rangle \left\langle n^{j}\right\rangle }{\sqrt{\underset{k}{\sum}\left(n_{k}^{i}\right)^{2}-T\left\langle n^{i}\right\rangle ^{2}}\sqrt{\underset{k}{\sum}\left(n_{k}^{j}\right)^{2}-T\left\langle n^{j}\right\rangle ^{2}}}
\end{equation}

where $i$ and $j$ stand here for the different species to be considered and $k$ for the box index.

\subsection*{Marchenko--Pastur filtering}
Let us consider $N_s$ different species randomly distributed in a square grid with a total number of cells $T=n_x\cdot n_y$. If the condition $N_s\rightarrow\infty$, $T\rightarrow\infty$, but keeping $1<\frac{T}{N_s}<\infty$, is fulfilled, the distribution of eigenvalues of their correlation matrix follows the Marchenko--Pastur distribution\cite{Marchenko1967, Plerou1999, Laloux1999, PRX_Gar}, which reads,

\begin{equation}
 \rho(\lambda)=\frac{T}{N_s}\frac{\sqrt{(\lambda_+-\lambda)(\lambda-\lambda_-)}}{2\pi\lambda},~~\mathrm{if}~~\lambda_-\leq\lambda\leq\lambda_+,
\end{equation}

and $\rho(\lambda)=0$ otherwise. The extreme eigenvalues take the value $\lambda_\pm=(1\pm\sqrt{\frac{N_s}{T}})^2$. Thus, any eigenvalue falling within this range can be considered as caused by noise, while eigenvalues out of this range represent meaningful correlations.

We assume a correlation matrix that can be written in the following way\cite{PRX_Gar},
\begin{equation}
 C=C^{(r)}+C^{(s)}
 \label{eq:reconstruct}
\end{equation}

where $C^{(r)}$ represents the random matrix to be subtracted from the original correlation matrix, and $C^{(s)}$ the \emph{bona fide} correlation matrix between species.

In practice we diagonalize $C$ and rebuild the filtered matrix from the modes
that the Marchenko--Pastur law cannot account for,
\begin{equation}
 C^{(s)}=\underset{\lambda_i>\lambda_+}{\sum}\lambda_i\,\mathbf{v}_i\mathbf{v}_i^{T},
\end{equation}
$\mathbf{v}_i$ being the eigenvector of eigenvalue $\lambda_i$. Every mode
falling inside the noise band is discarded, so that the entries of $C^{(s)}$
that survive are the interspecific correlations that cannot be produced by
finite sampling alone. All filtered correlations quoted in the text are entries
of $C^{(s)}$ at the stated scale $\ell$.

For further discussion on the Marchenko-Pastur law, see Supplementary Note 2.

\subsection*{The neutrality scale}
The neutrality scale $\ell_N$ is read off the scale dependence of the number of
non-noise eigenvalues (Fig.~\ref{fig:panel15}d), comparing the mean over the
eight censuses with the mean and standard deviation of the neutral ensemble.
The approach is gradual: the empirical value first enters the band of two
standard deviations at 26~m, leaves it at 27--28~m, and stays within it from
29~m up to 45~m, the largest scale at which the Marchenko--Pastur law still
applies to the neutral domain (Supplementary Note 2). We therefore quote
$\ell_N$ as a range, 25--30~m, rather than as a single threshold.

\subsection*{Environmental regression}
To separate the associations mediated by
the measured environment from the rest, each species' occupation vector is
regressed on the soil variables mapped over the plot, and the correlation
matrix is recomputed on the residuals. Writing $n^i_k$ for the abundance of
species $i$ in cell $k$ and $e_{v,k}$ for the value of soil variable $v$ in
that same cell, we fit, for each species $i$, a model linear and quadratic
in each soil variable,

\begin{equation}
n^i_k = \beta^i_0 + \sum_v \left(\beta^i_v\, e_{v,k} + \gamma^i_v\, e_{v,k}^2\right) + \varepsilon^i_k,
\end{equation}

and take the residuals $\varepsilon^i_k$ as the part of the spatial
distribution of species $i$ that the measured environment does not explain.
The soil variables entering $e_{v,k}$ are available at the official website of ForestGEO \cite{soilmaps} and contain  estimated concentrations (mg/Kg) of base cations, extractable P, and ammonium and nitrate for the BCI 50 ha Forest Dynamics Plot at a scale of $\ell=20$~m. 

The residual correlation matrix is then built from the $\varepsilon^i$
exactly as $C$ is built from the $n^i$, and filtered in the same way, so
that the two matrices compared in Fig.~\ref{fig:soil}a differ only by the regression
step. A pair whose association is environmentally mediated has a smaller
residual correlation than the raw one; a pair whose environmental and
residual contributions have opposite signs has a larger one. Fig.~\ref{fig:soil} is
computed at $\ell = 20$~m, the resolution at which the soil data
are available.

\subsection*{Spatially explicit neutral model}
The neutral benchmark against which the empirical spectra are compared is a
spatially explicit voter model with speciation, in which all individuals are
demographically equivalent irrespective of the species they belong to. Each
site of a $500\times500$ periodic lattice carries exactly one individual, so
that the total abundance is conserved and only the species composition
evolves.

The dynamics proceeds through elementary death--replacement events (Fig.~\ref{fig:panel15}a). At each step a site is drawn uniformly at random
and its occupant is removed. With probability $\nu$ the vacancy is filled by
an individual belonging to a species not currently present in the system,
which is the speciation event; otherwise it is filled by a copy of an
individual drawn uniformly among those occupying the square of half-width $k$
sites centred on the vacancy. Dispersal is therefore local, and the kernel size $k$
is the only length built into the model. One generation corresponds to as
many elementary events as there are sites.

Stationary configurations are sampled exactly, with no burn-in, through the
dual process of coalescing random walks\cite{Villegas2021, Pigolotti2018}.
The two parameters were fixed at $\nu=9\times10^{-5}$ and $k=3$ sites.
These values lie in the region of
parameter space that Villegas et al. identified as
reproducing the BCI rank-abundance curve with the same model and algorithm. Species richness was not calibrated independently: it
emerges from the same $(\nu,k)$ pair, giving a mean of 285 species
per realization against $\sim300$ in BCI. The same studies showed that, for
local kernels, this model also reproduces the short-range clumping of the pair
correlation function and the anomalous density fluctuations observed in BCI,
which provides validation on spatial statistics independent of the
abundance distribution; we did not repeat those tests here.

Each stationary configuration is then processed exactly as an empirical
census: it is partitioned into cells of side $\ell$, restricted to its 100 most abundant
species, and turned into a correlation matrix whose spectrum is compared with
the Marchenko--Pastur prediction calibrated for its system size. Averaging over 100 independent
realizations yields the neutral ensemble whose mean and standard deviation
define the bands of Fig.~\ref{fig:panel15}c,d. Because the empirical
and the simulated data pass through an identical coarse-graining and
filtering pipeline, any residual difference between the two spectra cannot be
attributed to either of those two steps.

\subsection*{Conspecific pair probability}
For each census, stems were placed on a 1-m raster of the plot and the number
of pairs of stems at distance $r$ was obtained from the spatial
autocorrelation of the raster of each species, computed by fast Fourier
transform with zero padding so that the plot is not wrapped. $F(r)$ is the
number of conspecific pairs at distance $r$ divided by the number of all pairs
at the same distance, which corrects for the plot edges without further
assumptions. Distances are binned linearly up to 10~m and logarithmically
above. The logarithmic law is fitted by least squares for $r\geq25$~m. In
Fig.~\ref{fig:pairs}b the numerator is restricted to the 112 species with at
least 200 stems, and the conspecific pairs of the $k$ most abundant of them
are removed from it; the denominator always counts all pairs.

\subsection*{Neutral model for the pair probability}
The neutral curve of Fig.~\ref{fig:pairs}b uses the same voter model with
speciation and square kernel, sampled through its dual, on a
$1360\times1360$ periodic lattice whose central $680\times340$ window is read
as the plot. The lattice spacing is then 1.47~m, which reproduces the density
of stems above 1~cm dbh in BCI. The kernel half-width, $k=14$ sites
(20.6~m), was set so that the slope of $F(r)$ for $r\geq25$~m matches the
empirical one (0.0044 against 0.0042), and $\nu=2\times10^{-5}$ so that the
window holds about 280 species (about 300 in BCI). The model does not
reproduce the dominance of the two most abundant species of BCI, so its
$F(r)$ lies below the empirical one at all distances (0.024 against 0.062 at
25~m), and Fig.~\ref{fig:pairs}b compares shapes only. Three independent
realizations were used.

\section*{Acknowledgements}
P.V. acknowledges the Spanish Ministry and Agencia Estatal de Investigaci\'on (AEI), MICIN/AEI/10.13039/501100011033, for financial support, Project PID2023-149174NB-I00 funded also by ERDF/EU. T.G. acknowledges the EU project ``HumanE-AI-Net,'' no.~952026.

The authors acknowledge Jim Dalling, Robert John, Kyle Harms, Robert Stallard and Joe Yavitt for their work in collecting and analyzing soil data (see the section on data availability).
\newpage

\makeatletter\let\auto@bib@innerbib\relax\makeatother

\onecolumngrid
\clearpage
\pdfximage{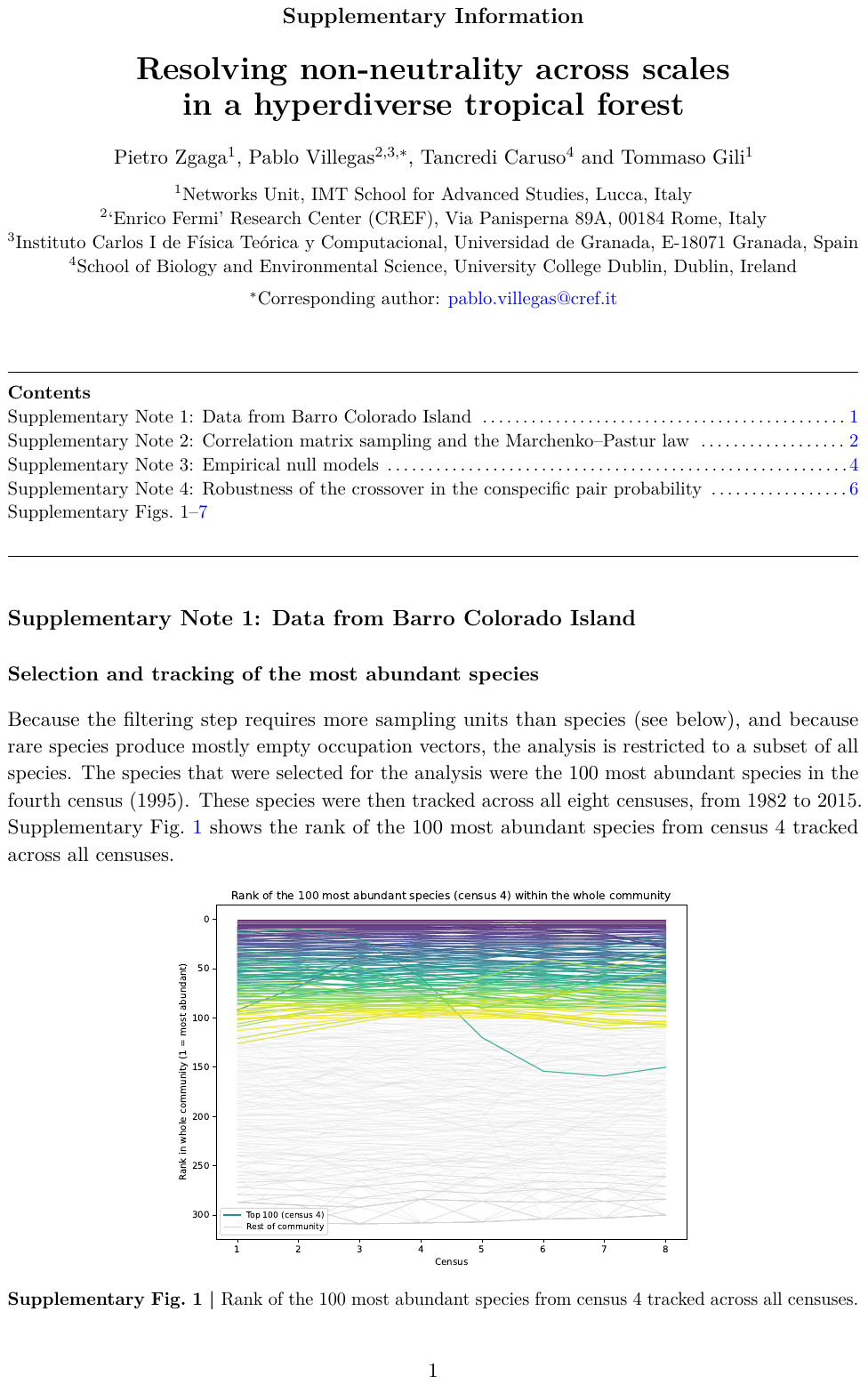}\edef\SIpages{\the\pdflastximagepages}
\newcount\SIfirst \SIfirst=\getpagerefnumber{SIfirst}\relax
\AddToShipoutPictureFG{%
  \ifnum\SIfirst>0 \ifnum\value{page}<\SIfirst\else
    \AtPageLowerLeft{\textcolor{white}{\rule{\paperwidth}{\paperheight}}}
    \AtPageLowerLeft{\includegraphics[width=\paperwidth,height=\paperheight,page=\numexpr\value{page}-\SIfirst+1\relax]{SI.pdf}}%
  \fi\fi}
\null\label{SIfirst}\clearpage
\foreach \p in {2,...,\SIpages}{\null\clearpage}

\end{document}